\documentstyle[11pt]{article}
\font\sm=cmr9
\newcommand {\be}{\begin{equation}} 
\newcommand{\fe}{\end{equation}}
\newcommand{\eqn}{\label}

\begin{document}

\title{REGULARISATION OF CLASSICAL SELF INTERACTION IN STRINGS}

\author{ {\bf Brandon Carter}
\\ D.A.R.C., (UPR 176, CNRS),
\\ Observatoire de Paris, 92 Meudon, France.}

\date{20 December, 1997}

\maketitle

{\bf Abstract:} A general method of regularisation of 
classical self interaction in strings is extended from the electromagnetic
case (for which it was originally developed) to the gravitation case,
for which the result can also be represented as a renormalisation.

\section{Introduction }

Concentration on supersymmetric quantum string theory has diverted attention
from basic problems that remain to be dealt with in classical string theory.
However even while laying the foundations of quantum electrodynamics, Dirac
found time to obtain a regularised Lorentz covariant treatment of a classical
electromagnetically self interacting point particle.  Following this example,
my purpose here is to treat the analogous problem of self 
interaction in classical relativistic string 
models. 

I shall consider only the kinds of long range interaction
that are experimentally familiar, namely electromagnetic interactions as
governed by Maxwell's equations for $A_\mu$, and weak gravitational
interactions as governed by the linearised Einstein equations for a small
amplitude but perhaps rapidly varying perturbation $h_{\mu\nu}=\delta
g_{\mu\nu}$ of a slowly varing spacetime metric $g_{\mu\nu}$ characterising a
4-dimensional background with local coordinates $x^\mu$. In a suitable 
gauge, the field equations reduce to the standard forms
\be \nabla{\!_\sigma}\nabla^\sigma A^\mu=-4\pi\hat j^\mu\, ,\eqn{1}\fe
and
\be \nabla_{\!\sigma}\nabla^\sigma h^{\mu\nu}=-8\pi 
\hbox{\sm G} \big(2\hat T^{\mu\nu}
-\hat T_{\!\sigma}^{\,\sigma}g^{\mu\nu}\big)\, , \eqn{2}\fe
in which $\hat j^{\mu}$ is the electromagnetic current density, 
$\hat T^{\mu\nu}$ is the stress momentum energy density tensor, and 
{\sm G} is Newton's constant. 

The problem of ultraviolet divergences for point particle or string models
arises because, in these cases, the relevant source densities $\hat j^\mu$ and
$\hat T^{\mu\nu}$ are not regular functions: they will be Dirac type
distributions that vanish outside the relevant one or two dimensional
worldsheets. In the case of a string with local worldsheet coordinates
$\sigma^a$ ($a=0,1$) and induced metric $\gamma_{ab}=g_{\mu\nu}
x^\mu{_{\! ,a}} x^\mu{_{\! ,b}}$, 
the latter will be expressible using the terminology of Dirac delta
``functions'' in the form
\be\hat T^{\mu\nu}=\Vert g\Vert^{-1/2}\int \overline T{^{\mu\nu}}\, 
\delta^{\rm 4}[x-x\{\sigma\}]\, \Vert\gamma \Vert^{1/2}
\, d^2\sigma\, ,\eqn{4}\fe
and there will be a similar relation between $\hat j{^\mu}$ and 
$\overline j{^\mu}$, where the surface stress momentum energy density 
$\overline T{^{\mu\nu}}$,  and the surface current $\overline j{^\mu}$ are
{\it regular} tensorial functions on the worldsheet (but undefined off it). 
The string case is more awkward than that of a point particle, since is beset
by infrared as well as ultraviolet divergences. This complication has so far
prevented the construction of a satisfactorily Lorentz covariant string
analogue of Dirac's (finite) radiation reaction formula for the point particle
case. 

\section{Regularisation}

What {\it can} be done for the string case is the analogue of the familiar 
treatment of the dominant (lowest differential order) contribution, which is 
divergent, and needs to be regularised by a cut-off procedure, following which 
-- in the point particle case -- it turns out to be absorbable by a simple mass
renormalisation. It will be seen that an analogous, but not so simple,
renormalisation is also possible in the string case. The basic force balance
equation will be expressible in the form
\be\overline\nabla_{\!\nu}\overline T{^{\mu\nu}}= 
f_{\rm e}{^\mu}+ f_{\rm g}{^\mu}\, ,\eqn{5}\fe
in which 
the relevant tangentially projected differentiation operator is defined by
$\overline\nabla_{\!\nu}=\eta_\nu{^\mu}\nabla_{\!\mu}$ where
$\eta^{\mu\nu}=\gamma^{ab}x^\mu{_{\!,a}}x^\nu{_{\!,b}}$ is the 
fundamental tensor of the worldsheet, 
while the electromagnetic force density
contribution will be given by the familiar formula
\be f_{\rm e}{^\mu}=F{^\mu}{_\nu}\overline j{^\nu}
\, , \hskip 1 cm F_{\mu\nu}=\nabla_{\![\mu}
A_{\nu]}\, ,\eqn{6}\fe
and, by the results of a recent gravitational perturbation
analysis\cite{BC95}, the gravitational force contribution will be given in
terms of the relevant worldsheet supported hyper-Cauchy tensor $\overline{\cal
C}{^{\mu\nu\rho\sigma}}$ by an expression of the form
\be f_{\rm g}{^\mu}= {_1\over^2}\overline T{^{\nu\sigma}}
\nabla^\mu h_{\nu\sigma} - \overline\nabla_{\!\nu}\big(
\overline T{^{\nu\sigma}} h_\sigma{^\mu}+
\overline{\cal C}{^{\mu\nu\rho\sigma}} h_{\rho\sigma} \big)
\, . \eqn{7}\fe
These force densities would evidently be well behaved if the fields $A_\mu$
and $h_{\mu\nu}$  were due just to passing radiation. However we are concerned
with the case in which they are obtained as the appropriate Dalembertian Green
function solutions of the source equations (\ref{1}), (\ref{2}) , which give
values that (while finite outside) are divergent on the worldsheet where they
are needed. 

As in the point particle case, one can obtain realistically regularised
values, $\widehat A_\mu$ and $\widehat h_{\mu\nu}$ say,  on the string
worldsheet itself, by taking account of the fact that the physical system one
wishes to describe (a vacuum vortex defect in the cosmic 
string case) will not really be 
infinitely thin but will have a finite thickness that provides an
appropriately microscopic ``ultraviolet'' cut-off length, $\delta_\ast$ say.
In the string case it is also be necessary to introduce a long range
``infrared'' cut-off length, $\Delta$ say, that might represent the
macroscopic mean distance between neighbouring strings. The relevant Green
function integrals will then be proportional to a logarithmic regularisation
factor of the form $ \widehat l= {\rm ln}\big\{ {\Delta^2 /\delta_\ast^{\, 2}
} \big\}$. More specifically (as pointed out in his original discussion 
of ``superconducting'' cosmic strings by Witten) the dominant contribution to
the regularised electromagnetic self field $\widehat A_\mu$ on the string
will be obtained~\cite{C97} in the simple form 
\be \widehat A_\mu=\widehat l\
\overline j_\mu\, .\eqn{12}\fe 
By comparing (\ref{1}) and (\ref{2}), it can thus be seen that the 
corresponding expression for the regularised gravitational self field,
$\widehat h_{\mu\nu}$ say, will have the form 
\be \widehat h_{\mu\nu}=2\hbox{\sm G}\,\widehat l\,\big(2\overline T_{\!\mu\nu}
- \overline T_{\!\sigma}{^{\!\sigma}}g_{\mu\nu}\big)\, .\eqn{13}\fe
(If the microscopic electromagnetic source distribution were very different
from that of the gravitational source distribution, the natural cut-off
$\delta_\ast$ that would be most appropriate for the former might be somewhat
different from what would be most appropriate for the latter, but since the
dependence on the cut off is only logarithmic there will not usually be any
significant loss of accuracy in using the same regularisation factor $\widehat
l$ for both cases.)

For substitution in (\ref{6}) and (\ref{7}) -- to get correspondingly
regularised self force contributions $\widehat f_{\rm e}^{\,\mu}$ and
$\widehat f_{\rm g}^{\,\mu}$ -- knowledge of the simple regularised self
fields $\widehat A_\mu$ and $\widehat h_{\mu\nu}$ is not sufficient. These
regularised values are well defined only on the worldsheet and so do not
provide what is needed for a direct evaluation of the gradients that are
required: there is no difficulty for the terms involving just the tangentially
projected gradient operator $\overline\nabla_{\!\nu}$, but there are also
contributions from the unprojected gradient operator $\nabla_{\!\nu}$, which
is directly meaningfull only when acting on fields whose support extends off
the worldsheet. 

\section{The regularised gradient operator}

It fortunately turns out that that this problem has a very simple and elegant
solution~\cite{C97}. The appropriate regularisation 
of the gradients on the string worldsheet turns out to be obtainable simply 
by replacing the ill defined operator $\nabla_{\!\nu}$ by a corresponding 
regularised gradient operator that is given in terms of the worldsheet 
curvature vector,
$K^\mu=\overline\nabla_{\!\nu}\eta^{\mu\nu}$, by the formula 
\be\widehat\nabla_{\!\nu}=\overline\nabla_{\!\nu}+{_1\over ^2}K_\nu \,
.\eqn{14}\fe 
For the appropriately regularised electromagnetic field tensor on the
string worldsheet this gives 
\be \widehat F_{\mu\nu}=2\widehat\nabla_{\![\mu}\widehat A_{\nu]} =\widehat
l\,\big(2\overline\nabla_{\![\mu}\overline j_{\nu]} +K_{[\mu}\overline
j_{\nu]}\big) \, ,\eqn{15}\fe 
which, by the surface current conservation condition $\overline\nabla_{\!\nu}
\overline j{^\nu}=0$, implies that the corresponding electromagnetic self
force contribution in (\ref{6}) will be expressible in the form
\be \widehat f_{\rm e}^{\,\mu}=-\overline\nabla_{\!\nu} \widehat T_{\!\rm
e}{^{\mu\nu}} \, ,\eqn{16}\fe 
where the relevant stress momentum energy density contribution from the
electromagnetic self interaction is 
\be \widehat T_{\!\rm e}{^{\mu\nu}}=\widehat A{^\mu} \overline j{^\mu}
-{_1\over^2}\widehat A_\rho\overline j{^\rho}\eta^{\mu\nu} \, .\eqn{17}\fe
It can be seen from (\ref{7}) that the regularised gravitational self force 
contribution will be similarly expressible in the form
 \be \widehat f_{\rm g}^{\,\mu}=-\overline\nabla_{\!\nu} \widehat T_{\!\rm
g}{^{\mu\nu}} \, ,\eqn{18}\fe 
in which the relevant self gravitational stress momentum energy density
contribution works out as
\be \widehat T_{\!\rm g}{^{\mu\nu}}=\widehat h_\sigma{^\mu} \overline
T{^{\nu\sigma}} -{_1\over^4}\widehat h_{\rho\sigma}\overline
T{^{\rho\sigma}}\eta^{\mu\nu} +\widehat h_{\rho\sigma}\overline{\cal
C}{^{\rho\sigma\mu\nu}} \, .\eqn{19}\fe

It evidently  follows that the self force contributions will be absorbable by
a renormalisation whereby the original ``bare'' 
stress momentum energy density tensor $\overline T{^{\mu\nu}}$ is replaced by 
the ``dressed'' stress momentum energy density tensor 
\be\widetilde T{^{\mu\nu}}=\overline T{^{\mu\nu}}+
\widehat T_{\rm e}{^{\mu\nu}}+\widehat T_{\rm g}{^{\mu\nu}}\, ,\eqn{20}\fe
so that -- in the absence of radiation from outside -- the basic force 
balance equation (\ref{5}) will reduce to a surface energy momentum 
conservation law of the simple form 
$\overline\nabla_{\!\nu}\widetilde T{^{\mu\nu}}=0$.

\end{document}